\documentclass[
	aps, reprint,
	amsmath,amssymb,
	superscriptaddress,
	pra
   ]{revtex4-1}

\usepackage{graphicx}
\usepackage{color}
\usepackage{dcolumn}
\usepackage{siunitx}
\usepackage{bm}
\usepackage{braket}
\begin{document}


\title[]{Quadratic Clifford expansion for efficient benchmarking and initialization of variational quantum algorithms}

\author{Kosuke Mitarai}
\email{mitarai@qc.ee.es.osaka-u.ac.jp}
\affiliation{Graduate School of Engineering Science, Osaka University, 1-3 Machikaneyama, Toyonaka, Osaka 560-8531, Japan.}
\affiliation{Center for Quantum Information and Quantum Biology, Institute for Open and Transdisciplinary Research Initiatives, Osaka University, Japan.}
\affiliation{JST, PRESTO, 4-1-8 Honcho, Kawaguchi, Saitama 332-0012, Japan.}

\author{Yasunari Suzuki}
\affiliation{NTT Secure Platform Laboratories, NTT Corporation, Musashino 180-8585, Japan}
\affiliation{JST, PRESTO, 4-1-8 Honcho, Kawaguchi, Saitama 332-0012, Japan.}

\author{Wataru Mizukami}
\affiliation{Center for Quantum Information and Quantum Biology, Institute for Open and Transdisciplinary Research Initiatives, Osaka University, Japan.}
\affiliation{JST, PRESTO, 4-1-8 Honcho, Kawaguchi, Saitama 332-0012, Japan.}

\author{Yuya O. Nakagawa}
\affiliation{QunaSys Inc., Aqua Hakusan Building 9F, 1-13-7 Hakusan, Bunkyo, Tokyo 113-0001, Japan}

\author{Keisuke Fujii}
\affiliation{Graduate School of Engineering Science, Osaka University, 1-3 Machikaneyama, Toyonaka, Osaka 560-8531, Japan.}
\affiliation{Center for Quantum Information and Quantum Biology, Institute for Open and Transdisciplinary Research Initiatives, Osaka University, Japan.}
\affiliation{Center for Emergent Matter Science, RIKEN, Wako Saitama 351-0198, Japan}

\date{\today}

\begin{abstract}
    
Variational quantum algorithms are considered to be appealing applications of near-term quantum computers.
However, it has been unclear whether they can outperform classical algorithms or not.
To reveal their limitations, we must seek a technique to benchmark them on large scale problems.
Here, we propose a perturbative approach for efficient benchmarking of variational quantum algorithms.
The proposed technique performs perturbative expansion of a circuit consisting of Clifford and Pauli rotation gates, which is enabled by exploiting the classical simulatability of Clifford circuits.
Our method can be applied to a wide family of parameterized quantum circuits consisting of Clifford gates and single-qubit rotation gates.
The approximate optimal parameter obtained by the method can also serve as an initial guess for further optimizations on a quantum device, which can potentially solve the so-called ``barren-plateau'' problem.
As the first application of the method, we perform a benchmark of so-called hardware-efficient-type ansatzes when they are applied to the VQE of one-dimensional hydrogen chains up to $\mathrm{H}_{24}$, which corresponds to $48$-qubit system, using a standard workstation.

\end{abstract}

\pacs{Valid PACS appear here}
\maketitle

\section{Introduction}

As promising candidates for possible applications of early-days quantum devices, variational quantum algorithms (VQAs) have been developed rapidly.
VQAs utilize parameterized quantum circuits $U(\bm{\theta})$ with parameters $\bm{\theta}$ which are optimized with respect to some suitably defined cost function $\mathcal{L}(\bm{\theta})$ depending on specific tasks.
Since there are quantum circuits that cannot be simulated classically \cite{Arute2019}, we might gain practical speedups from VQAs if we choose $U(\bm{\theta})$ from such ones.
This choice of $U(\bm{\theta})$ guarantees classical inability to perform exactly the same tasks.

Target applications of VQAs range among quantum chemistry calculations \cite{Peruzzo2014,McClean2016,Kandala2017}, combinational optimization \cite{farhi2014quantum} and machine learning \cite{Mitarai2018, Schuld2019,Havlicek2019,farhi2018classification,benedetti2019generative}.
Despite the vast amount of theoretical proposals, demonstrations and benchmarks of algorithms, whether they be experiments with actual quantum devices \cite{arute2020quantum, Arute2020hartree, Nam2020, Havlicek2019} or numerical simulations, are limited to relatively small scale problems, where classical simulations are still feasible.
Efficient techniques for their benchmark in large scale problems are strongly demanded to understand the limitations of VQAs and to develop more sophisticated algorithms.

Here we aim to resolve the above problem with a perturbative expansion of the cost function.
We assume that a parametric circuit utilized in an algorithm is made of Clifford gates and single-qubit rotation gates whose angles are the circuit parameters and initialized to zero. The above form of the circuit includes a wide family of parameterized circuits, which guarantees the wide applicability of the proposed method.
With the assumption, we can efficiently compute first and second derivatives of a cost function by exploiting the classical simulatability \cite{10.5555/1972505, Aaronson2004} of Clifford circuits.
This allows us to perform a simple minimization of a quadratic function to obtain an approximately optimal value of parameters and a cost function.
In particular, for the variational quantum eigensolver (VQE) \cite{Peruzzo2014}, which is an algorithm to obtain an approximate ground state of a quantum system, the perturbative treatment can be justified because classically tractable variational solutions such as Hartree-Fock or mean-field states are expected to be close to true ones.
Our method serves as a performance benchmark of a VQA in such cases.

The method can also be seen as an efficient initializer of the circuit parameters as the above procedure corresponds to the first step of Newton-Raphson optimization.
The obtained Hessian together with the obtained parameter can be passed as an initial guess to quasi-Newton optimizer such as the BFGS method for further optimization on quantum devices.
This method can potentially solve the so-called ``barren-plateau'' problem \cite{McClean2018} which states that the gradient of the cost function vanishes exponentially to the number of qubits when circuits and parameters are chosen randomly.
The problem itself is not an obstacle to the proposed method, because the whole process is run on a classical computer where the accuracy of the computation can be improved exponentially with only polynomially scaling resource, which is in contrast to the case of using quantum devices.

In the following, we first give the concrete algorithm of the proposed method.
Then, as an application of the method, we benchmark the so-called hardware-efficient-type
ansatzes applied to the VQE up to 48 qubits using the hydrogen chain as a testbed.
This benchmark can be performed because the proposed technique provides us an approximately optimal value of the cost function, which is the energy expectation value in this case.
The benchmark itself can be seen as a demonstration of a ``quantum-inspired'' quantum chemistry calculation, where the ansatz wavefunction is constructed with the language of the quantum circuit.
Finally, we also show the effectiveness of our perturbative initialization approach by another numerical experiment.

\section{Algorithms}

\subsection{Variational quantum algorithms}

VQA generally refers to a family of algorithms that involve use of parametrized quantum circuits $U(\bm{\theta})$ whose parameters $\bm{\theta}$ are optimized with respect to a suitable cost function $\mathcal{L}(\bm{\theta})$.
$U(\bm{\theta})$ is used to generate a $n$-qubit parametrized state $\ket{\psi(\bm{\theta})}:=U(\bm{\theta})\ket{0}^{\otimes n}$.
Hereafter we abbreviate $\ket{0}^{\otimes n}$ by $\ket{0}$ when it is clear from the context.
A famous example of VQAs is the VQE \cite{Peruzzo2014}, where the cost function is defined as an energy expectation value $E(\bm{\theta})$ with respect to a Hamiltonian $H$, i.e. $\mathcal{L}(\bm{\theta})=E(\bm{\theta})=\bra{\psi(\bm{\theta})}H\ket{\psi(\bm{\theta})}$.
The cost function is usually computed from expectation values of observables also in other examples such as machine learning \cite{Mitarai2018, Schuld2019,benedetti2019generative,farhi2018classification, Havlicek2019}, and combinational optimization \cite{farhi2014quantum}.
A general form of $\mathcal{L}(\bm{\theta})$ can be written as $\mathcal{L}(\bm{\theta}) = L\left(\braket{O(\bm{\theta})}\right)$ where $O$ denote the measured observable and $\braket{O(\bm{\theta})}:= \bra{\psi(\bm{\theta})}O \ket{\psi(\bm{\theta})}$.
$O$ is typically expressed as a sum of $N_o$ $n$-qubit Pauli operators $\{P_i\} \subset \{I,X,Y,Z\}^{\otimes n}$ as $O=\sum_{i=1}^{N_o} c_i P_i$ with coefficients $\{c_i\}$.

\subsection{Main result: Quadratic Clifford expansion}

We consider an ansatz in the form of
\begin{align}
    U(\bm{\theta}) = R_K(\theta_K)C_K \cdots R_2(\theta_2)C_2 R_1(\theta_1) C_1 \label{eq:ansatz},
\end{align}
where $\bm{\theta}=\{\theta_k\}_{k=1}^K$, $R_k$ is a single-qubit rotation gate generated by a Pauli operator $P_k$, i.e. $R_k(\theta_k)=e^{i\theta_k P_k}$, and $C_k$ is a circuit consisting of Clifford gates.
We furthermore assume $R_k(0)$ to be Clifford.
Note that this form of the ansatz is quite general.
When we wish to build a hardware-efficient ansatz \cite{Kandala2017}, it is frequently in the form of Eq. (\ref{eq:ansatz}) because the two-qubit gates which are tuned to give a high-fidelity on the hardware are usually Clifford gates such as controlled-NOT or controlled-Z gates.
More sophisticated ansatz such as unitary coupled-cluster (UCC) \cite{Peruzzo2014} can also be written in this form.

This form of the ansatz allows us to efficiently compute the perturbative form of the cost function $\mathcal{L}(\bm{\theta})$. 
More concretely, a Taylor expansion of $\braket{O(\bm{\theta})}$ around $\bm{\theta}=0$ can be written as,
\begin{align}
    \braket{O(\bm{\theta})} 
    &=\braket{O(0)} + \sum_k  g_k\theta_k + \frac{1}{2}\sum_{k,m} A_{km}\theta_k \theta_m + \mathcal{O}(\|\bm{\theta}\|^3),\label{eq:taylor}
\end{align}
where,
\begin{align}
    g_k &= 2\mathrm{Re}\left[\bra{0} U^\dagger(0) O \frac{\partial U(0)}{\partial \theta_k}\ket{0}\right], \\
    \begin{split}
        A_{km} &= 2\mathrm{Re}\left[\bra{0} \frac{\partial U^\dagger(0)}{\partial \theta_k} O \frac{\partial U(0)}{\partial \theta_m}\ket{0}\right] \\
        &\quad+ 2\mathrm{Re}\left[\bra{0} U^\dagger(0) O \frac{\partial }{\partial \theta_k} \frac{\partial U(0)}{\partial \theta_m}\ket{0}\right],
    \end{split}
\end{align}
and $\frac{\partial U(0)}{\partial \theta_k}:=\left.\frac{\partial U(\bm{\theta})}{\partial \theta_k}\right|_{\bm{\theta}=0}$, which can then be used to expand $\mathcal{L}(\bm{\theta})=L(\braket{O(\bm{\theta})})$ itself.
Since we assumed $U$ to be in the form of Eq. (\ref{eq:ansatz}), $\frac{\partial U(0)}{\partial \theta_i}$ can be efficiently computed.
To see this, observe that,
\begin{align}
    \frac{\partial U(0)}{\partial \theta_k} = iR_K(0)C_K\cdots P_k R_k(0)C_k \cdots R_2(0)C_2 R_1(0) C_1.
\end{align}
Since we assume $R_k(0)$ to be Clifford for all $k$, $P_k$ can be passed through $R_K(0)C_K \cdots R_{k+1}(0)C_{k+1}$.
Let 
\begin{align}
    \begin{split}
    &R_K(0)C_K \cdots R_{k+1}(0)C_{k+1} P_k \\
    &= P_k'R_K(0)C_K \cdots R_{k+1}(0)C_{k+1}
    \end{split}
\end{align}    
for some Pauli operator $P_k'$.
$P_k'$ can be found in time $\mathcal{O}(nK)$ on a classical computer if $\{C_k\}$ are local.
Using $P_k'$, the coefficients appearing in the second-order Taylor expansion (Eq. (\ref{eq:taylor})) can be written in the terms of the expectation values $\braket{\psi(0)|OP_k'|\psi(0)}$, $\braket{\psi(0)|P_k'OP_m'|\psi(0)}$ and $\braket{\psi(0)|OP_k'P_m'|\psi(0)}$.
The decomposition of the operators $OP_k'$, $P_k'OP_m'$ and $OP_k'P_m'$ into a sum of Pauli operators can be computed in $\mathcal{O}(nN_o)$ on a classical computer.
This can be performed simply by multiplying $P_k'$ and $P_m'$ to each Pauli operator in $O$. 
The expectation values of these operators can also be evaluated classically because $\ket{\psi(0)}$ is a stabilizer state under the assumption that $U(0)$ is Clifford.
More concretely, we evaluate expectation values of each Pauli operator constituing $OP_k'$, $P_k'OP_m'$ and $OP_k'P_m'$, and then take the summation.
This process can be performed in time $\mathcal{O}(n^2N_oK^2)$ using a standard simulation technique \cite{Aaronson2004}, which gives the leading order complexity of the perturbative expansion.
We call this technique quadratic Clifford expansion.
The technique itself might be useful for classical simulations of near-Clifford circuits.

The perturbative expansion given in Eq. (\ref{eq:taylor}) is justified especially for the VQE \cite{Peruzzo2014}, where we can obtain an approximate ground state classically by using techniques such as Hartree-Fock methods.
If we construct $U(\bm{\theta})$ in such a way that $\ket{\psi(0)}$ becomes the Hartree-Fock ground state, $\ket{\psi(0)}$ is considered to be close to the true ground state.
Therefore, in this case, we can presume the optimal value of $\bm{\theta}$ to be small, which justifies the perturbative treatment of the cost function $E(\bm{\theta})$.
As long as the perturbation is accurate enough, we can obtain the optimal value of $\braket{O(\bm{\theta})}$ by simply minimizing the quadratic function obtained with the second-order expansion, which can be done in time $\mathcal{O}(K^3)$ and provides us an optimal parameter $\bm{\theta}^*=-A^{+}\bm{g}$, where $A^{+}$ is the Moore-Penrose pseudo-inverse of the Hessian $A$.
The approximate, perturbative optimal value of $\braket{O(\bm{\theta})}$ can be calculated by substituting $\bm{\theta}^*$ into Eq. (\ref{eq:taylor}) and neglecting cubic error term, which we denote by $\braket{O}^{*}$.

It must be noted that whether the above perturbation is accurate or not cannot be determined classically in general, since we cannot obtain the expectation value $\braket{O(\bm{\theta})}$ for general $\bm{\theta}$ on a classical computer.
This indicates the need for a quantum device for validating the result.
Therefore, a possible strategy of using the proposed technique is to evaluate $\braket{O(\bm{\theta}^*)}$ on a quantum computer, and if it returns a value close to the perturbative one, then we just conclude that we have found an optimal approximate ground state; otherwise, we further optimize the parameters starting from $\bm{\theta}^*$.
Note that $\bm{\theta}^*$ can also be seen as the parameter obtained by the first step of the Newton-Raphson method and provides a good starting point for further optimization.
Moreover, the Hessian $A_{km}$ can be passed to quasi-Newton optimizers such as the BFGS method which are frequently utilized in the VQE.

We can also apply the proposed method to the machine learning algorithms \cite{Mitarai2018, Schuld2019,benedetti2019generative,farhi2018classification, Havlicek2019}
In this direction, techniques for computing a ``good'' initial guess like mean-field solution are not yet developed to the best of our knowledge.
However, for example, we might be able to obtain an initial guess by using the correspondence of the Boltzmann machine and the quantum circuit developed in \cite{PhysRevResearch.2.033125}.
Also, one can use the gradient and Hessian of the cost function obtained by the above protocol for performing the first optimization step and also pass the latter to the quasi-Newton optimizers.

\section{Numerical experiment}

To demonstrate the effectiveness of our idea, we apply the method described in the previous section to the VQE to benchmark the performance of so-called hardware-efficient ansatz \cite{Kandala2017}.
For this purpose, we use electronic Hamiltonians of evenly-spaced one-dimensional chains of hydrogen atoms $\mathrm{H}_m$, which are frequently used as a benchmark system for quantum chemistry calculations~\cite{PhysRevX.7.031059}.
All benchmarks are performed on a workstation with two Intel Xeon Silver 4108 processors.
For quantum circuit simulations, we utilized an NVIDIA Tesla-V100 GPU.

\subsection{Experimental details}

Electronic Hamiltonians of hydrogen chains are generated by OpenFermion \cite{McClean_2020} and PySCF \cite{sun2018pyscf,sun2020recent} using the STO-3G minimal basis set.
The generated fermionic Hamiltonians are mapped to qubit ones by Jordan-Wigner transformation implemented in OpenFermion, which results in a $2m$-qubit Hamiltonian for an $m$-hydrogen chain $\mathrm{H}_m$.
A thorough review of these procedures can be found at e.g. Ref. \cite{RevModPhys.92.015003}.
All conventional quantum chemistry calculations are also performed with PySCF.

\begin{figure}
    \centering
    \includegraphics[width=\linewidth]{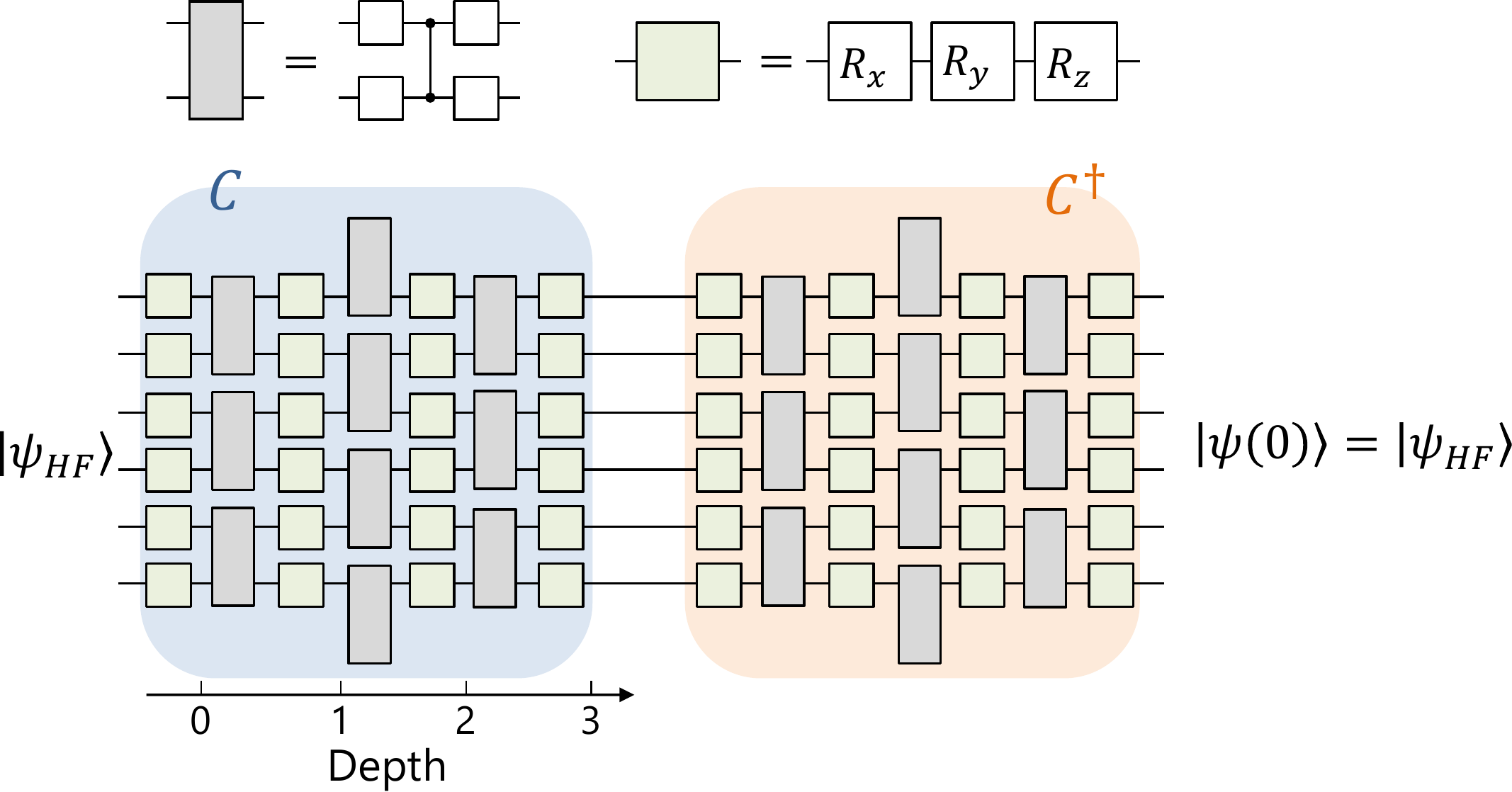}
    \caption{Ansatz used in the numerical experiment. Grey boxes represent (fixed) 2-qubit Clifford gates in the form of the upper left, where white boxes are randomly chosen from 24 single-qubit Clifford gates \cite{Selinger_2015}. Green boxes represent parametrized single qubit rotations consisting of $x$, $y$ and $z$-axis rotations respectively.}
    \label{fig:ansatz}
\end{figure}

As for the ansatz, we use the one shown in Fig. \ref{fig:ansatz}, which can be regarded as a ``hardware-efficient'' ansatz constructed on a one-dimensional qubit array.
It consists of alternating layers of two-qubit Clifford gates and single-qubit rotation gates.
This form of the ansatz can generate sufficiently non-local evolutions that give non-zero gradients.
In Fig. \ref{fig:ansatz}, the two-qubit Clifford gates in the region shaded by blue are randomly chosen as shown in the upper left of the figure.
Those in the orange region are chosen so that the overall circuit becomes identity when $\bm{\theta}=0$.
This allows us to easily guarantee $\ket{\psi(0)}$ to be the Hartree-Fock state $\ket{\psi_{HF}}$; we can just inject $\ket{\psi_{HF}}$ to the input of the circuit.
Note that Hartree-Fock states are computational basis states under fermion-to-qubit mappings such as Jordan-Wigner transformation, and its evolution under Clifford gates can efficiently be simulated.
The single-qubit rotations hold the parameter $\bm{\theta}$ to be optimized.
Each of them has three parameters as $x$, $y$, and $z$-rotation angles.
Note that, although the two-qubit Clifford gates are chosen to satisfy the above constraint, the parameters implemented in single-qubit rotations in each region are independent.

Using this ansatz, we calculate the gradient and Hessian based on the method described in the previous section.
Then, we perform the minimization of the second-order perturbative energy to obtain approximately optimal energies and parameters.
This provides us $\bm{\theta}^*$ and perturbative energies $E^*=\braket{H}^*$.
Finally, when possible, we simulate an ansatz whose parameters are set to the perturbatively optimal ones to check if the perturbative treatment can be justified.
This simulation is performed with Qulacs \cite{Qulacs}. 

To make the benchmark systematic, we set the depth of the ansatzes equal to the number of hydrogen atoms, $m$.
This scaling of the depth can be considered as the largest possible value for today's most advanced quantum computer \cite{Arute2019}.
Note that this choice corresponds to $L=\mathcal{O}(m^2)$.
Combining with the fact that $N_o = \mathcal{O}(m^4)$ and $n=2m$ in this case, the overall time complexity of computing Hessian for this system is $\mathcal{O}(m^8)$.
Since the minimization of the quadratic function can be done in time $\mathcal{O}(L^3)=\mathcal{O}(m^6)$, the Hessian part contributes the most to the total time.

Finally, to somewhat relax the randomness of the ansatz, we first randomly generate 200 ansatzes in the form of Fig.~\ref{fig:ansatz} for each $m$ used in the experiment.
For each generated ansatz, we calculate $g_l$ using the Hamiltonian with the spacing of 1.0 \AA.
Then, the circuit with the largest $\sum_l |g_l|$ is chosen as the ansatz to be used for each hydrogen chain with different spacings.
This is based on our expectation that an ansatz with large gradients would provide the highest performance.

\subsection{Results and discussion}

\subsubsection{Benchmark results of small-scale systems}
\begin{figure}
    \centering
    \includegraphics[width=\linewidth]{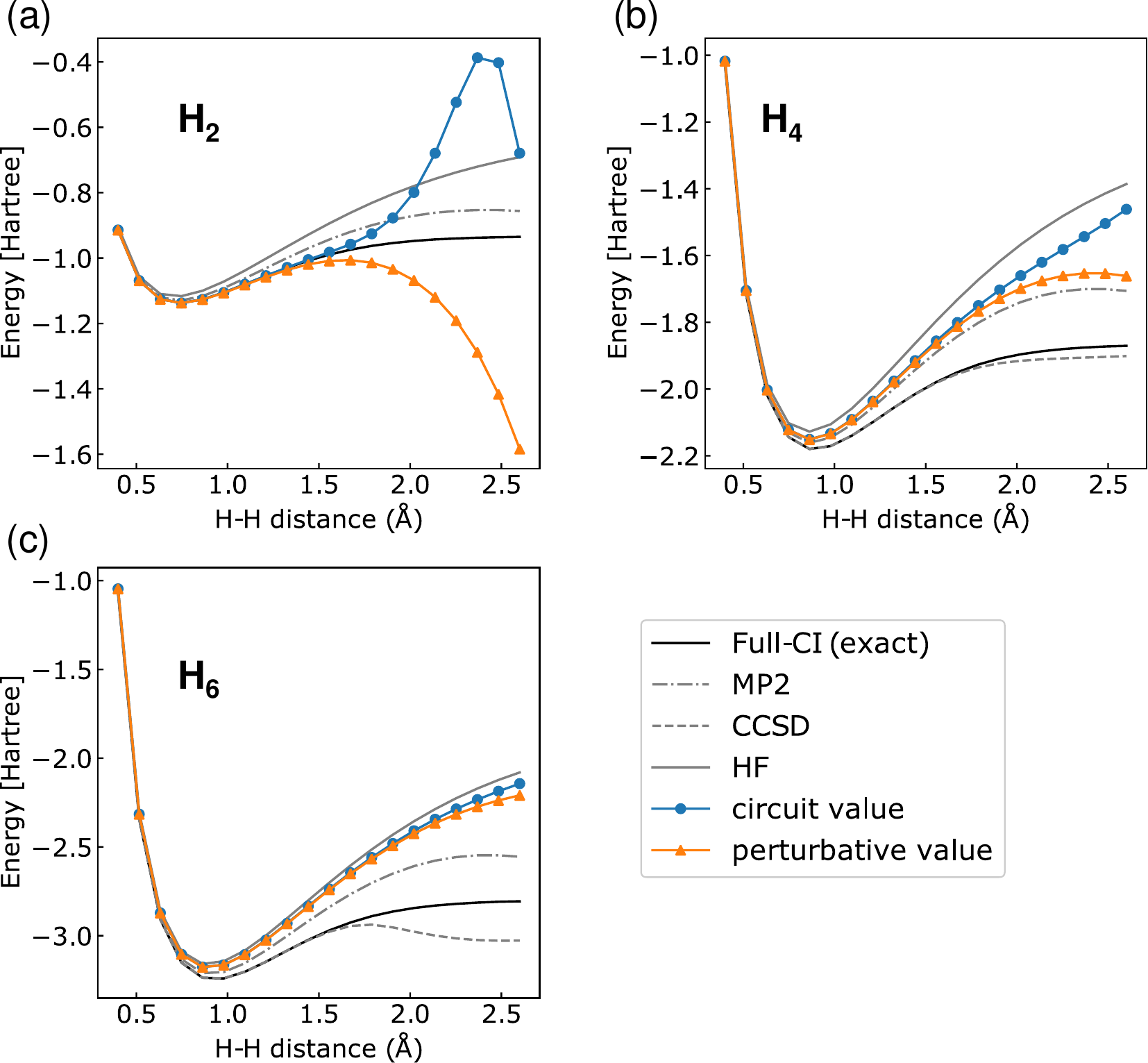}
    \caption{Results of the numerical experiments at (a) $m=2$, (b) $m=4$ and (c) $m=6$. Full-CI is the exact ground state energy, and MP2, CCSD, and HF respectively mean approximate ground state energy obtained with second-order M{\o}ller-Plesset perturbation, coupled cluster with single and double excitations, and Hartree-Fock. Circuit value and perturbative value represent $E(\bm{\theta}^*)$ and its perturbative approximation computed from Eq. (\ref{eq:taylor}), respectively.}
    \label{fig:SmallScaleResults}
\end{figure}

Figure \ref{fig:SmallScaleResults} shows the result of the numerical experiment at $m=2$, $4$, and $6$ along with the energy obtained from standard quantum chemistry calculations as references.
For $m=2$ and $4$, we can observe that the energies obtained from the circuit simulation and the one from the perturbative optimization match well at small spacings.
Here, the Hartree-Fock method gives a relatively accurate description of the ground state, and the perturbative treatment works fine as expected.
The effectiveness of the perturbation also means that we can achieve the optimal parameter with this technique.
This implies that the hardware-efficient ansatzes considered in this work can only achieve the accuracy of second-order M{\o}ller-Plesset perturbation (MP2), which is a technique used widely in current quantum chemistry calculations as one of the easiest post-Hartree-Fock methods, for $\mathrm{H}_4$ as we can observe from Fig. \ref{fig:SmallScaleResults} (b).
There is a possibility of improving the accuracy by optimizing from a randomly initialized $\bm{\theta}$ as the above discussion only considers the case where we take $\bm{\theta}=0$ as the initial parameter. 
However, such a strategy would not be generally scalable because of the barren plateau problem \cite{McClean2018}.
On the other hand, the perturbative treatment breaks down at the larger spacings where the electronic correlation becomes larger.
As mentioned in the previous section, one has to perform further optimization in such a case.

In the case of $m=6$, we cannot observe the clear break down of the perturbative treatment, i.e., the energies obtained from the perturbation match well with the ones from the circuit simulation.
Again, it means that the optimal parameters and corresponding energies can be obtained with the proposed technique.
We can see that the hardware-efficient circuit cannot even achieve the MP2 energy for $\mathrm{H}_6$.
Note that MP2 considers up to double electron excitations and involves $\mathcal{O}(n^4)$ parameters in its construction. 
This scaling is considerably greater than the number of parameters implemented in the ansatz of Fig. \ref{fig:ansatz} with depth $n$.
In this sense, the performance worse than MP2 is expected behavior.
This trend of decreasing accuracy will also be certified with the result in the next subsection.

\subsubsection{Benchmark results of large-scale systems}\label{sec:largescaleresults}

Although the complexity of $\mathcal{O}(m^8)$ is polynomial to $m$, the large exponent prohibits us from extending the analysis to larger scales.
Due to this complexity, we need some modifications to the experimental settings.
First, we modify the ansatz to only involve real numbers by generating an ansatz that has the same form as the one shown in Fig. \ref{fig:ansatz}, but the random single-qubit Clifford gates are randomly chosen from identity and Hadamard gates only, and the single-qubit rotations just contain a $y$-rotation.
This modification reduces the number of parameters by a third.
We find that this modification does not significantly alter the result as shown in Appendix, which can be explained by the fact that the eigenstates of non-relativistic quantum chemistry Hamiltonians can be described with states that are real in the computational basis.
Second, to further reduce the number of parameters involved in the Hessian calculation, we ``drop out'' the parameters that give zero gradients to the energy, i.e., $y$-rotations that do not give the first-order contribution to the energy are removed from the ansatz after calculating the gradient.
This is motivated by our observation in the preliminary experiment where we have found that the gradients with respect to most of the parameters are exactly zero.
In the following experiment, the results are obtained by removing the rotation gates with $|g_l|< 10^{-6}$ Hartree.
In Appendix, we show that this modification does not significantly alter the results either.

\begin{figure}
    \centering
    \includegraphics[width=\linewidth]{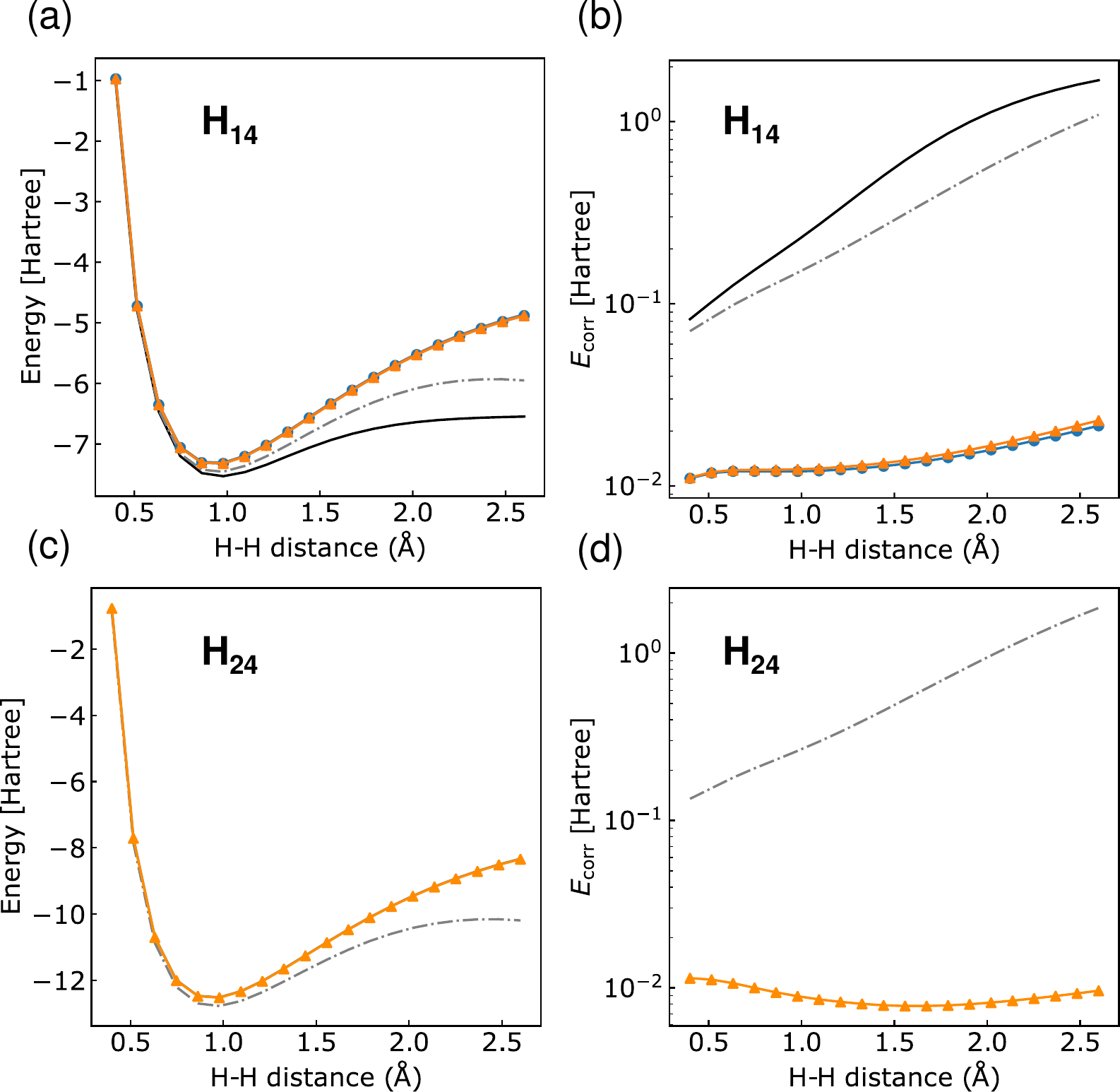}
    \caption{Results of the numerical experiments at (a), (b) $m=14$ and (c), (d) $m=24$. Graph legends follow those of Fig. \ref{fig:SmallScaleResults}. (a) and (c) show the total energy, while (b) and (d) show the correlation energy.}
    \label{fig:LargeScaleResults}
\end{figure}

Figure \ref{fig:LargeScaleResults} shows the results for $m=14$ and $m=24$ cases.
Note that $m=14$ corresponds to 28 qubits, which is the largest number of qubits that can be handled with the Qulacs-GPU simulator \cite{Qulacs} using an NVIDIA Tesla-V100 processor.
At $m=16$, the required memory far exceeds its capacity.
For this reason, we do not show the circuit value in the case of $m=24$ which corresponds to 48 qubits (Fig. \ref{fig:LargeScaleResults} (c), (d)).
This is, to the best of our knowledge, the first benchmark of the VQE at this scale.
Also, as the exact diagonalization at $m=24$ could not be performed under our environment, it is not shown in the figure.
The energy of coupled-cluster with single and double excitations (CCSD) is not depicted in both cases because we experienced its numerical instability.

In Fig. \ref{fig:LargeScaleResults} (b) and (d), we show the correlation energy $E_{\mathrm{corr}}$ defined as the difference between the obtained energy and the Hartree-Fock reference energy to illustrate the performance of the proposed method and the hardware-efficient ansatz itself.
From Fig. \ref{fig:LargeScaleResults} (b), we can observe that the perturbative energy and the circuit value match almost exactly, indicating that we can analyze the performance of the ansatz itself with this perturbative treatment.
Therefore, Fig. \ref{fig:LargeScaleResults} (b) and (d) show that the correlation energy that can be achieved by the hardware-efficient ansatz used in this work is a tenth or a hundredth smaller than that of MP2.
Note that the perturbative minimum obtained by this method is a local one that locates around $\bm{\theta}=0$, and the global optimal solution for this ansatz can perform better.
Nevertheless, this result indicates the need for more structured ansatzes such as unitary coupled-cluster \cite{Peruzzo2014} to correctly capture the correlation of electrons in a molecule.

\subsubsection{Initialization performance}

Finally, we show that the approximate optimal parameter $\bm{\theta}^*=-A^+ \bm{g}$ together with the initial Hessian $A$ can indeed serve as a good initial guess of the VQE.
To this end, we take Hamiltonians of $\mathrm{H}_4$ at different atom spacings as examples, and compare the convergence of the optimization procedure of the VQE when using different initial parameters $\bm{\theta}_{\mathrm{init}}$.
The BFGS method implemented in SciPy \cite{virtanen2020scipy} which is a popular quasi-Newton technique is employed as the optimizer.
We compare three cases: $\bm{\theta}_{\mathrm{init}}=0$, $\bm{\theta}_{\mathrm{init}}=-A^{+}\bm{g}$, and $\bm{\theta}_{\mathrm{init}}=-A^{+}\bm{g}$ with the initial Hessian provided to the optimizer.

\begin{figure}
    \centering
    \includegraphics[width=0.8\linewidth]{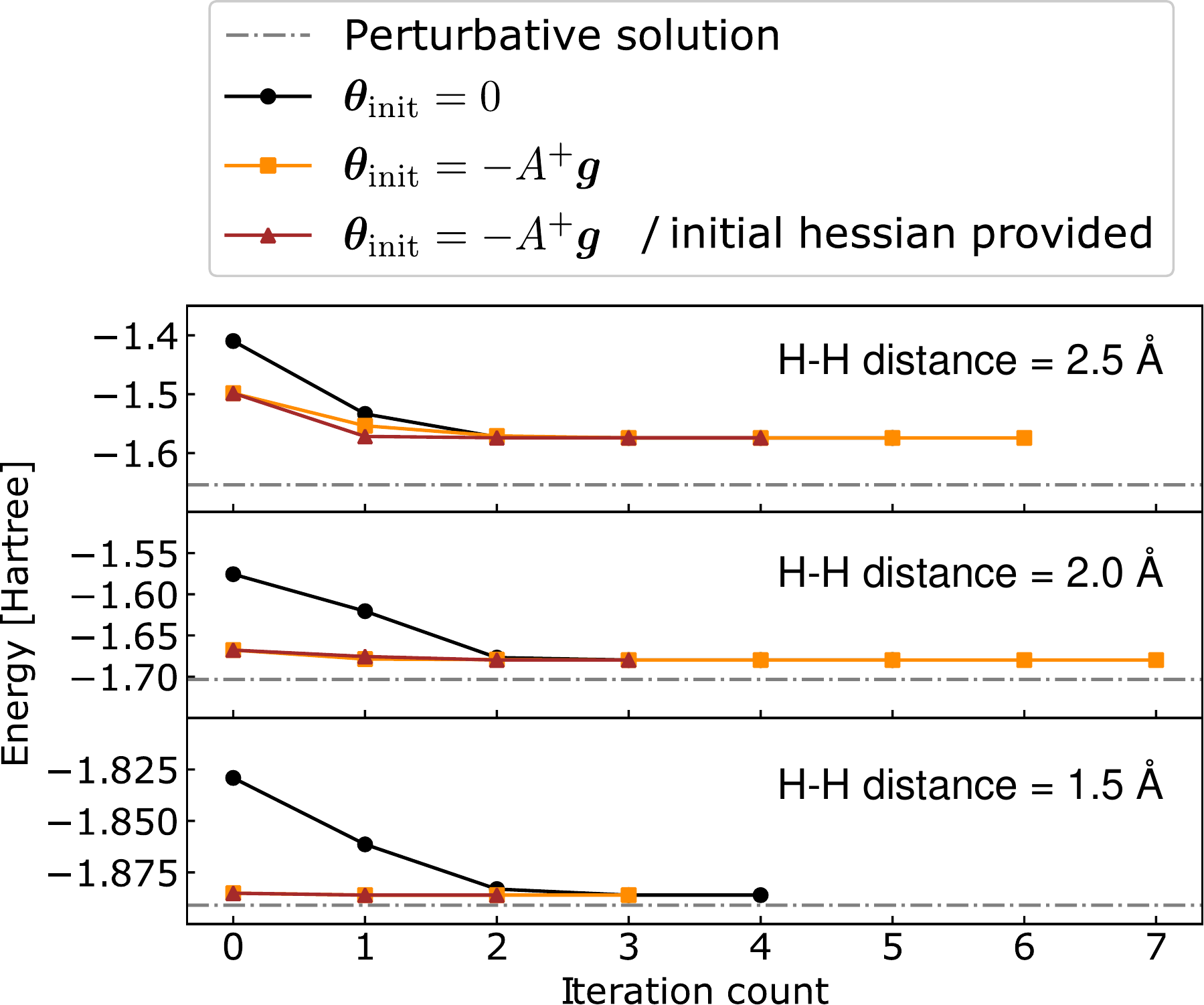}
    \caption{Comparison of the convergence of optimization procedures with different $\bm{\theta}_{\mathrm{init}}$ using $m=4$ Hamiltonian at different spacings of hydrogen atoms.}
    \label{fig:VQE}
\end{figure}

Figure \ref{fig:VQE} shows the result of the numerical experiment.
We can observe that the $\bm{\theta}_{\mathrm{init}}=-A^{+}\bm{g}$ cases exhibits the faster convergence than $\bm{\theta}_{\mathrm{init}}=0$ in all cases.
Also, the optimizer with the initial Hessian performs equally well to or better than the case $\bm{\theta}_{\mathrm{init}}=-A^{+}\bm{g}$ without providing Hessian.
This result demonstrates the effectiveness of our perturbative initialization approach.

\section{Conclusion}

We proposed a technique to efficiently compute an approximate optimal parameter and the corresponding value of the cost function in the VQAs.
It is based on the observation that we can efficiently compute the gradient and Hessian of the cost function if an ansatz is in the form of Eq. (\ref{eq:ansatz}) which includes a wide range of circuits.
Since the method is based on a perturbative expansion, we can obtain an accurate solution when the initial guess of the parameter from which we perform the Taylor expansion of Eq. (\ref{eq:taylor}) is close to an optimal one.
Even if we do not have such an initial guess, the gradient and Hessian can be used to perform the first step optimization, and those quantities can be passed to optimizers.
The generality of the ansatz allows us to apply the proposed method to various VQAs such as VQE \cite{Peruzzo2014,McClean2016,Kandala2017}, quantum approximate optimization \cite{farhi2014quantum}, and variational machine learning algorithms \cite{Mitarai2018, Schuld2019,Havlicek2019,farhi2018classification,benedetti2019generative}.

We applied the method to the VQE of hydrogen chains with a one-dimensional hardware-efficient ansatz shown in Fig. \ref{fig:ansatz} for its benchmark.
The numerical experiments showed that the performance of such a hardware-efficient ansatz in the VQE cannot even achieve that of classical MP2 calculation.
To the best of our knowledge, the proposed method is the only one that enables us the benchmark of the VQAs beyond the scale that is classically simulatable.
Although the benchmark results are pessimistic, it also motivates us to construct more structured ansatzes such as unitary coupled-cluster \cite{Peruzzo2014} and to make other initialization strategies such as the one presented in Ref. \cite{Grant2019initialization}.
For example, one might be able to use genetic optimization to improve the ansatz in Fig. \ref{fig:ansatz} from the random choice of Clifford gates.
One might also be able to improve the performance of this ansatz by using localized orbitals instead of the naive Hartree-Fock orbitals utilized in this work to express the Hamiltonian, which would make it easier for the ansatz to capture the electronic correlation.
We believe that the proposed technique will be of use to a wide range of the VQAs.

\begin{acknowledgments}
	KM is supported by JST PRESTO Grant No. JPMJPR2019 and JSPS KAKENHI Grant No. 20K22330. 
	YS is supported by JST PRESTO Grant No. JPMJPR1916.
	WM is supported by JST PRESTO Grant No. JPMJPR191A.
	KF is supported by JSPS KAKENHI Grant No. 16H02211,  JST ERATO JPMJER1601, and JST CREST JPMJCR1673.
	This work is supported by MEXT Quantum Leap Flagship Program (MEXT QLEAP) Grant Number JPMXS0118067394 and JPMXS0120319794.
\end{acknowledgments}

\appendix*

\section{Reducing the number of parameters in the ansatz}

\begin{figure}[h]
    \centering
    \includegraphics[width=\linewidth]{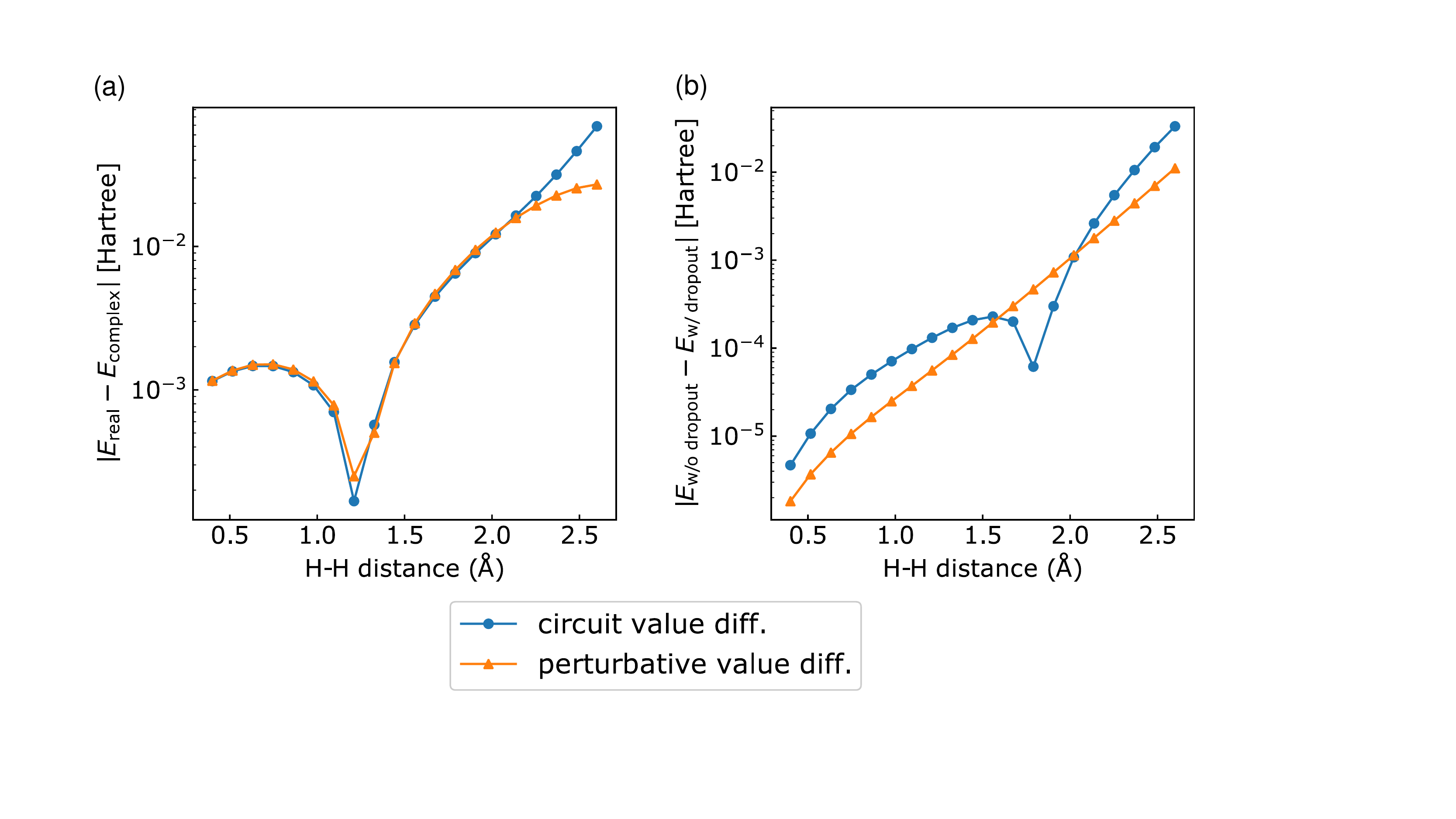}
    \caption{(a) Difference of the results obtained with the original ansatz in Fig. \ref{fig:ansatz} and the one modified to generate only real-valued state vectors at $m=6$. (b) Difference of the result with and without dropout at $m=6$ using the modified ansatz similar to Fig. \ref{fig:ansatz}, which only generates real wavefunctions. }
    \label{fig:dropout-compare}
\end{figure}

First, we demonstrate that the modification of the ansatz to be real in the computational basis (see Sec. \ref{sec:largescaleresults}) does not alter the results significantly.
It is illustrated in Fig. \ref{fig:dropout-compare} (a), where we plot the difference of the energy obtained by the modified and original ansatz, respectively denoted as $E_{\mathrm{real}}$ and $E_{\mathrm{complex}}$.
The energy difference does not exceed $10^{-2}$ Hartree in the figure when the spacing is less than 2.0 \AA, which is negligible compared to the correlation energy.

The effect of the ``drop-out'' utilized in Sec. \ref{sec:largescaleresults} does not alter results either.
Figure \ref{fig:dropout-compare} (b) shows the comparison of the results with and without dropout at $m=6$ using the modified real ansatz.
We can observe that the ``drop-out'' only slightly alters the result by about the same magnitude as Fig. \ref{fig:dropout-compare} (a). 

\end{document}